\documentclass[useAMS,usenatbib]{mn2e}

\usepackage{graphicx}
\usepackage{amsmath}

\usepackage{natbib}
\title[Area of accretion curtains in EX Hya]{On the area of accretion curtains from fast aperiodic time variability of the intermediate polar EX Hya}
\author[Semena et al.]{Andrey N. Semena$^{1}$\thanks{E-mail:
san@hea.iki.rssi.ru}, Mikhail G. Revnivtsev$^{1}$, David A.H. Buckley$^{2,3}$, 
\newauthor
Marissa M. Kotze$^{2}$, Ildar I. Khabibullin$^{1}$, Hannes Breytenbach$^{2,4}$
\newauthor
Amanda A. S. Gulbis$^{2}$, Rocco Coppejans$^{5}$, Stephen B. Potter$^{2}$\\
$^{1}$Space Research Institute, Russian Academy of Sciences, Profsoyuznaya 84/32, 117997 Moscow, Russia\\
$^{2}$ South African Astronomical Observatory, PO Box 9, 7935 Observatory, Cape Town, South Africa\\
$^{3}$ Southern African Large Telescope Foundation, PO Box 9, 7935 Observatory Cape Town, South Africa\\
$^{4}$ University of Cape Town, Private Bag X3, 7701 Rondebosch, South Africa\\
$^{5}$ Department of Astrophysics/IMAPP, Radboud University Nijmegen, P.O. Box 9010, 6500 GL Nijmegen, The Netherlands
}
\begin{document}

\date{}

\pagerange{\pageref{firstpage}--\pageref{lastpage}} \pubyear{2013}

\maketitle

\label{firstpage}

\begin{abstract}
We present results of a study of the fast timing variability of the magnetic cataclysmic variable (mCV) EX Hya. It was previously shown that one may expect the rapid flux variability of mCVs to be smeared out at timescales shorter than the cooling time of hot plasma in the post shock region of the accretion curtain near the WD surface. Estimates of the cooling time and the mass accretion rate, thus provide us with a tool to measure the density of the post-shock plasma and the cross-sectional area of the accretion funnel at the WD surface. We have probed the high frequencies in the aperiodic noise of one of the brightest mCV EX Hya with the help of optical telescopes, namely SALT and the SAAO 1.9m telescope. We place upper limits on the plasma cooling timescale $\tau<$0.3 sec, on the fractional area of the accretion curtain footprint $f<1.6\times10^{-4}$, and a lower limit on the specific mass accretion rate $\dot{M}/A>3$ g sec$^{-1}$ cm$^{-2}$. We show that measurements of accretion column footprints via eclipse mapping highly overestimate their areas. We deduce  a value of $\Delta r/r \la 10^{-3}$ as an upper limit to the penetration depth of the accretion disc plasma at the boundary of the magnetosphere. 
\end{abstract}

\begin{keywords}
accretion, accretion discs, X-rays: binaries -- stars: individual: EX Hya 
\end{keywords}

\section{Introduction}
Confining a hot (millions K) plasma in a magnetically controlled volume is not only the goal of the thermonuclear fusion reactors in terrestrial laboratories, but also is a reality near magnetic and relativistic compact objects in binary systems. It was realised quite some time ago that the strong magnetic fields of accreting white dwarfs and neutron stars disrupt accretion flows at some distance from their surfaces to form an accretion column (or curtain) around their magnetic poles \citep{pringle72,lamb74}. However, very little is currently known  about the geometry of these columns or curtains. It is likely that the best currently available direct measurement of the size of an accretion column footprint on a white dwarf surface, was obtained several years ago via eclipse mapping of the polar FL Cet \citep{odonoghue06}. 

Indirectly, the geometry of the accretion column/curtain can be estimated from the density of the hot plasma which is heated by a standing shock wave near the white dwarf (WD) surface. The density determines the cooling time of the plasma, thus knowing this cooling time and the total mass accretion rate in the flow we can have a handle on the accretion column/curtain geometry. \cite{langer81} showed that if the hot post shock region is cooling only via the bremsstrahlung emission it should be thermally unstable, generating quasi-periodic oscillations (QPOs) of the shock (see also \citealt{chevalier82,imamura96}). It was expected that such oscillations should be observable as optical brightness variations (see e.g. \citealt{larsson95}). In later studies it was shown that there were mechanisms, like influence of cyclotron cooling, which can stabilize the post shock region \citep[e.g.][]{chanmugam85,wu94,wu99,wu00,saxton01}. Another possible mechanism which might make the post-shock region oscillations invisible in lightucrves is incoherent variations of different parts of the accretion column. If accretion flow is not uniform, but spreaded over a large number of quasi-independent filaments, their individual oscillations might be not in phase with each other. Therefore even if the hot plasma cooling time scale is similar in these filaments, their total emission will be sum of incoherent brightness oscillations of individual filament and will be strongly dumped.


More recently, it was proposed that in spite of the relative absence of QPOs in mCVs, one can still probe the cooling time of the accretion column plasma via the properties of their aperiodic variability  \citep{semena12}. 
It was proposed that aperiodic variability of the mass accretion rate in accretion columns, which, as we know from observations of accreting neutron stars, continues towards frequencies above 100 Hz \cite[see e.g.][]{jernigan00}, should not produce strong variability in the luminosity at timescales smaller than the cooling time of plasma in the hot post-shock region. All variability at higher frequencies should be smeared out for periods less than this cooling timescale, producing a break in the power density spectrum of the brightness variations. 

The search for such a break was done in optical light of a bright northern hemisphere intermediate polar (IP), LS Peg \citep{semena13}.  Choosing to study mCVs in optical light has certain advantages. The main gain is in the relative increase in the number of optical photons generated by the WD/accretion disc, due to X-ray illumination from the post-shock region.  Depending on the geometry and inclination of the system, this number can far exceed that of X-ray photons reaching us directly from the post-shock region. 

Unfortunately, due to the small level of aperiodic variations of light of LS Peg at high frequencies ($<0.2$\% at $f>0.1$ Hz), only an upper limit ($<10$ sec) on the cooling timescale was obtained.

In this paper we are furthering the search for the cooling time generated turnover frequency in the rapid variability of mCVs. Our target is one of the brightest accreting magnetic white dwarfs in the southern hemisphere, EX Hya.

\section{EX Hya}
	EX Hya is one of the closest and brightest intermediate polars, which makes it one of the best candidates for searching for cooling time features in the power density spectra of optical flux variations in magnetic CVs. Physical parameters of the system used in this study are listed in Table \ref{pars}.

The mass accretion rate onto WD in this binary system demonstrates stochastic variability, which we see through variations of its X-ray and optical flux \cite[e.g.][]{watson74,revnivtsev11}. The accretion flow in this binary occurs via an accretion disk, disrupted at distance approximately $2\times 10^9$ cm \citep{siegel89,hellier97,revnivtsev11}.  Assuming that this disruption is due to influence of a dipole magnetic field of the white dwarf we can obtain an approximate estimate its magnetic field from a simple formula of magnetospheric radius of accreting compact object \cite[e.g.][]{pringle72}: $R_{\rm in}\sim \mu^{4/7} (GM_{\rm WD})^{-1/7} \dot{M}^{-2/7}$, where $\mu$ is the magnetic moment of the compact object. Substituting parameters from Table \ref{pars} we obtain $\mu\sim 2.5\times10^{30}$ G cm$^{-3}$, which corresponds to magnetic field strength on WD surface $B\sim 7$ kG. This value is in agreement with upper limit $B<1$ MG, inferred from spectroscopic observations of the system in infrared spectral band \citep{harrison07}. Such low magnetic field of the WD means that the cyclotron cooling mechanism is not important for the dynamics of the post shock region of the accretion column, that the dominant cooling is the optically thin plasma bremsstrahlung emission \cite[see e.g.][]{lamb79,chanmugam85}, and  that electrons and ions in the accretion column/curtain has the same temperatures \citep[e.g.][]{imamura87}

\begin{table}
\caption{Parameters of EX Hya, used in the paper}
\label{pars}
\begin{tabular}{l|c|l}
Parameter&Value&Reference\\
\hline
Orbital period, h&1.63&\\
Mass WD, $M_{\rm WD}$&$0.79~M_\odot$& 1\\
Radius WD,$R_{\rm WD}$&$0.7 \times 10^{9}$ cm&1\\
Secondary radius, $R_2$&$1.04\times10^{10}$ cm&1\\
Inner disc radius, $R_{\rm in}$&$1.9\times10^{9}$cm&2,3\\
 WD magnetic field, $B$&$<1$ MG&4\\
                      &$\sim7-8$ kG& 5\\
Mass accretion rate, $\dot{M}$&$3 \times 10^{15}$g sec$^{-1}$&6\\
Binary separation, $a$& $4.68\times10^{10}$ cm& 1\\
Binary inclination, $i$& 77.8$^\circ$cm&7\\
Luminosity, $L_{\rm x}$& $2.6 \times 10^{32}$ erg/sec  &1\\
\hline
\end{tabular}

1 - \cite{beuermann08}, 2 - \cite{siegel89}, 3-- \cite{revnivtsev11}, 4 - \cite{harrison07}, 5 - assuming the magnetospheric radius equals to $R_{\rm in}$, 6 - from $L_{\rm x}\approx GM_{\rm WD}\dot{M}(1/R_{\rm WD} - 1/R_{\rm in})$, 7 - \cite{Hellier87}
\end{table}
	
	We used numerical integration techniques similar to \cite{wu94} in order to estimate the plasma cooling time and thus the accretion channel cross section. The WD mass and radius were fixed while the accretion channel cross section was varied to obtain the expected cooling time. To check different possibilities, we made use of the flat channel geometry of \cite{wu94} and the dipole channel geometry of \cite{Canalle05}(see equations37, 38). In both cases gravity of the white dwarf and bremsstrahlung cooling were taken into account.

We assume that the field lines of the magnetic dipole, which form the boundary of the accretion curtain, start at the inner edge of the accretion disc $R_{\rm in}$, and that matter from the surrounding area then settles onto these field lines. We also assume, for simplicity, that the magnetic dipole and the spin axis of the WD are co-aligned  (this is of course an oversimplification but sufficient for our  present estimates). We can then put an upper limit on the accretion curtain cross-sectional area, $A < \pi R_{c}^2 = \pi R_{\rm WD}^3/ R_{\rm in} \sim 5.6\times 10^{17}$ cm$^2$, where $R_{\rm WD}$ is the white dwarf radius and $R_{\rm in}$ the inner radius of the disc.

For cross sectional area $A = 2.8 \times 10^{17}$ cm$^2$ the corresponding cooling time is $\tau \approx 15$ sec, while for cross-sectional area $A = 0.01 ~4 \pi R_{\rm WD}^2 = 6 \times 10^{16}$ cm$^2$ the cooling time $\tau \approx 6$ sec. In case of pure bremsstrahlung cooling, we can therefore expect that source flux variability should be smeared out at frequencies above $>0.05-0.5$ Hz.

\section{Aperiodic variations of the EX Hya optical light}

\begin{figure}
\includegraphics[width=\columnwidth]{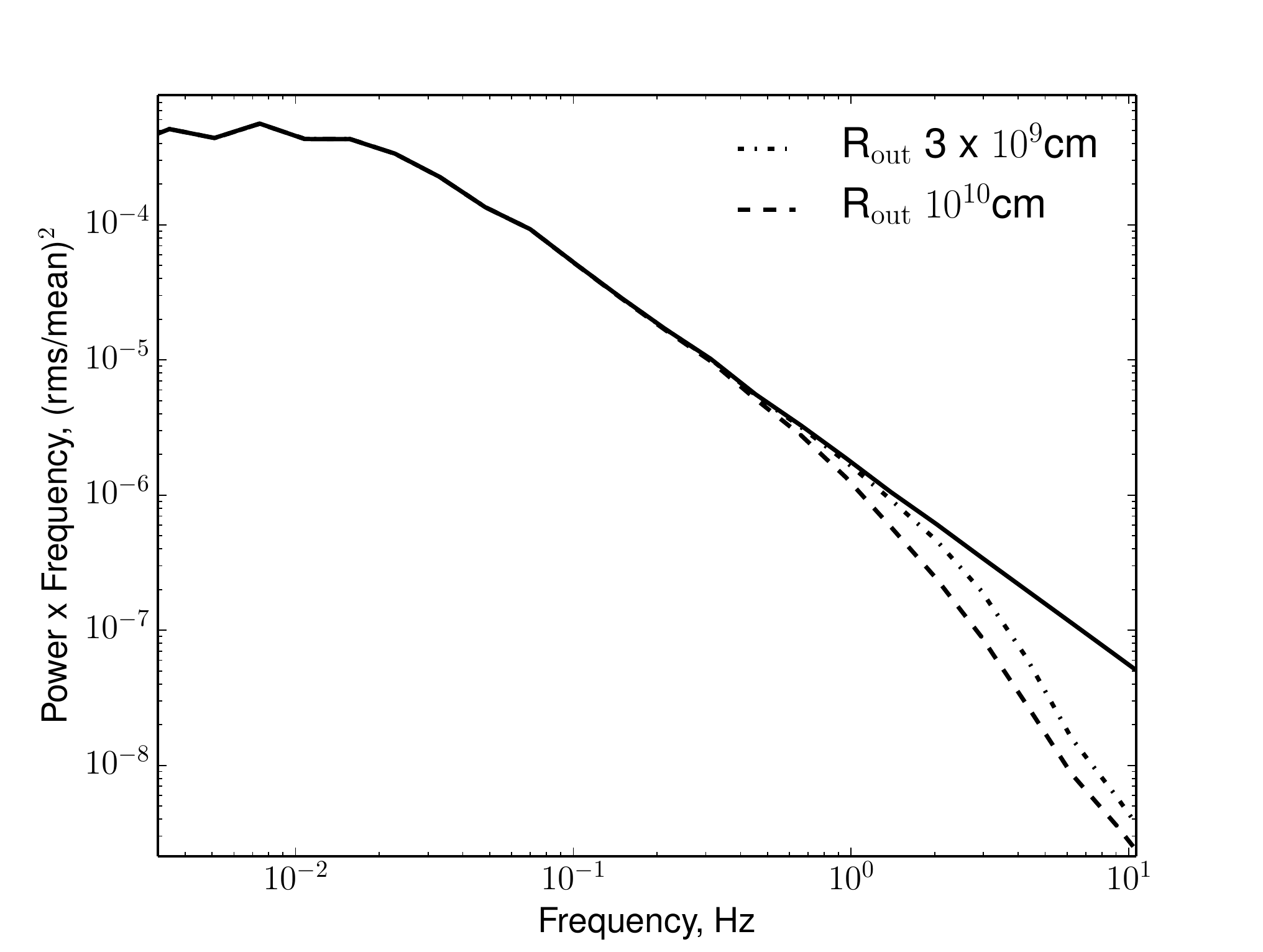}

\caption{Power spectrum of intrinsic variability of the source and its modifications due to smearing caused by reflection from disc. We present results of two models, where the reprocessor is an annulus with inner radius $R_{\rm in} = 1.9 \times 10^9$ cm  and the outer radius $r_{\rm out} = 3 \times 10^{9}$ cm (dotted curve) and $r_{\rm out} = 10^{10}$ cm (dashed curve)}

\label{transfer}
\end{figure}

Matter, which impacts the WD surface in accretion columns/curtains is heated up to high temperatures and emits X-ray radiation. The X-ray flux of even the brightest accreting magnetic WDs is relatively small ( on the order of $10^{-3}$  photons cm$^2$ sec$^{-1}$), which complicates any study of their fast timing variability.  However, part of this X-ray emission is intercepted by the underlying WD surface and the accretion disc and re-emitted in optical/UV energy range with much larger photon fluxes \citep{beuermann04,revnivtsev11}.  Any variations in X-ray luminosity of the post-shock region should therefore also be visible in optical light variations. These optical light variations look like flicker noise, often observed in accreting WD binaries \cite[see e.g.][]{bruch92,baptista04}. The goal of our paper is to use the optical light flickering to understand properties of the post-shock region.

Since we analyze the optical emission of the binary instead of its X-ray emission, we should take into account some modification of the emerging flux variability.

\subsection{Power Density Spectrum}
There are two main reasons for the power density spectrum of the optical flux variability to change: a) smearing of variability due to finite light crossing time of the reprocessor, and b) smearing, caused by a finite time, required for the reprocessor (inner part of the accretion disc of the WD surface) to absorb and re-emit incoming X-rays. The latter effect was considered e.g. by \cite{cominsky87}, where it was shown that the reprocessing timescale for parameters relevant for our study is not larger than $\sim1$ s (see also \citealt{hummer63,obrien02}).

In order to model the effect of smearing due to light crossing time of the reprocessing region we have calculated the transfer function of the reprocessor in two scenarios. First, we assumed that the reprocessor is an annulus with the inner disc radius $R_{\rm in} = 1.9 \times 10^9$ cm and: 1) the outer radius $R_{\rm out} = 3 \times 10^{9}$ cm, or 2) the outer radius $10^{10}$ cm.
Values of the outer radii of the reprocessing disc annulus (which is the most important parameter of the smearing kernel) were adopted using the following arguments. In the work of \cite{siegel89} it was shown that the optically bright region is located at a distance $\sim1.5\times10^{9}$ cm from the WD. The maximal outer radius of the optically bright region can be estimated also from the length of the optical eclipse: adopting the orbital inclination of the binary $i=77.8^\circ$ \citep{Hellier87} and the size of the secondary $R_2\sim1.04\times10^{10}$ cm \cite{beuermann08} the length of the eclipsing chord of the secondary can be estimated as $\la 5\times10^{9}$ cm. The shape of the eclipse shows that the emitting region should have a similar size.

In the model the illuminating X-ray source is raised at $0.6 \times 10^9$ cm above the plane of the disc, the disc has the shape $H\propto R^{9/8}$ (\cite{ss73}).

The power spectrum of simulated flux variability of an accreting magnetic WD, modified by disc reflection (along with power density spectrum of intrinsic variability) are shown in Fig.\ref{transfer}. It is seen that the effect of smearing in any case is important only at Fourier frequencies above few Hz.

\section{Observations}

In order to study properties of aperiodic variability of EX Hya in the optical, we have used a set of observations from telescopes at the South African Astronomical Observatory, including SALT. Table \ref{obslog} contains a list of observations.

\begin{table*}
\caption{Observations of EX Hya on telescopes of SAAO, used in this work.}
\begin{tabular}{c l c c c c}
\hline
Date & Time resolution, s & duration, s & UT start time & device& Filter\\
\hline
15.04.2010  & 1.0   & 5100 & 19:37 & HIPPO&R\\
17.04.2010  & 1.0   & 4700 & 23:21 & HIPPO&R\\
18.04.2010  & 1.0   & 4600 & 19:43 & HIPPO&R\\
29.03.2012  & 0.067 & 3239 & 02:52 & SHOC&White light\\
02.04.2012	& 0.067 & 3500 & 03:00 & SHOC&R\\
07.05.2012	& 0.1   & 2500 & 23:00	  & SALTICAM&CLR-S1\\	
07.02.2013  & 0.16  & 3760 & 23:50 & SHOC&White light\\
07.04.2013  & 0.1	& 1064 & 01:35 & BVIT&White light\\
02.08.2013  & 0.16  & 5118 & 00:53 & SHOC&White light\\
15.06.2013  & 0.057 & 7485 & 17:08 & SHOC& R\\
17.06.2013  & 0.032 & 7106 & 16:43 & SHOC& R\\
18.06.2013 & 0.032 & 3923 & 16:51 & SHOC& R\\
18.06.2013 & 0.032 & 4102 & 18:03 & SHOC& B\\
\hline
\end{tabular}
\label{obslog}
\end{table*}

\subsection{SALT/SALTICAM}

\begin{figure}
\includegraphics[width=\columnwidth]{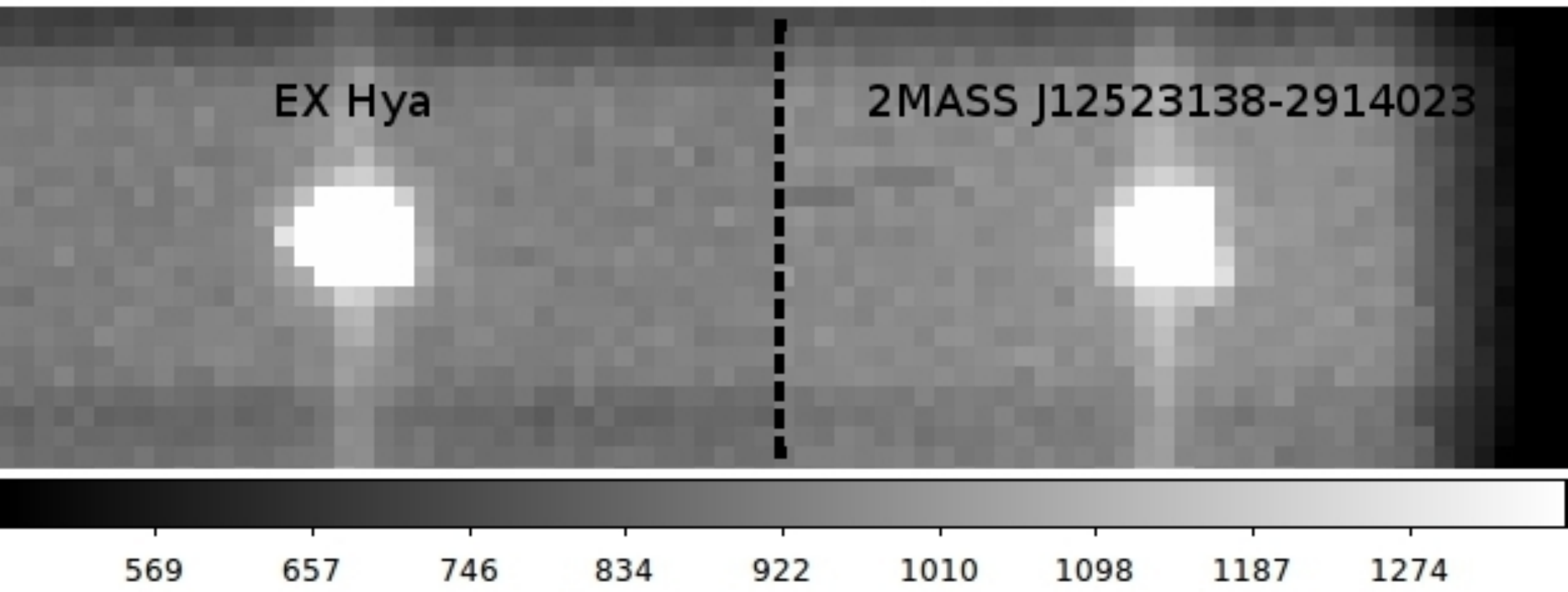}
\caption{Typical image frame with EX Hya (left object) and a comparison star (right object) collected over 0.1 sec. The grayscale shows the number of counts per pixel.}
\label{image}
\end{figure}

We obtained approximately 2.5 ks of fast photometry on the Southern African Large Telescope (SALT) using SALTICAM with a Sloan $r'$ filter on 7 May 2012, from 23:00 UTC to 24:00. Observations were performed in Slot Mode, with 6x6 binning, providing 0.1 s time resolution \citep{odonoghue06}. Slot Mode employs the following readout algorithm: a slot in front of the CCD leaves open only 144 unbinned rows (approximately 20 arcsec on the sky) of the CCD, just above the frame transfer boundary. During one exposure time, the signal is accumulated on the open part of the CCD. At the end of the exposure these 144 rows rapidly ($\sim 14$ msec) move over the frame transfer boundary. These images then migrate to the readout register in a stepwise manner. Start times of two consequent exposures differ by 104 ms (i.e. time resolution of our data). The SALTICAM readout algorithm has some features which affect data analysis. The time series have a time gap every 6.2 sec, lasting between $\sim$0.1 and $\sim$0.6 sec. Such 
periodic time gaps should be carefully taken into account in all studies of sources fluxes aperiodic variability power spectra. Therefore in our analysis we have split Fourier frequency range into two intervals, above and below 1/6.2 Hz. The power spectrum at Fourier frequencies above 1/6.2 Hz was obtained from evenly sampled data between gaps. The power spectrum at Fourier frequencies below 0.08 Hz was obtained from the lightcurve binned into 6.2 s time bins.

\begin{figure}

\includegraphics[width=\columnwidth]{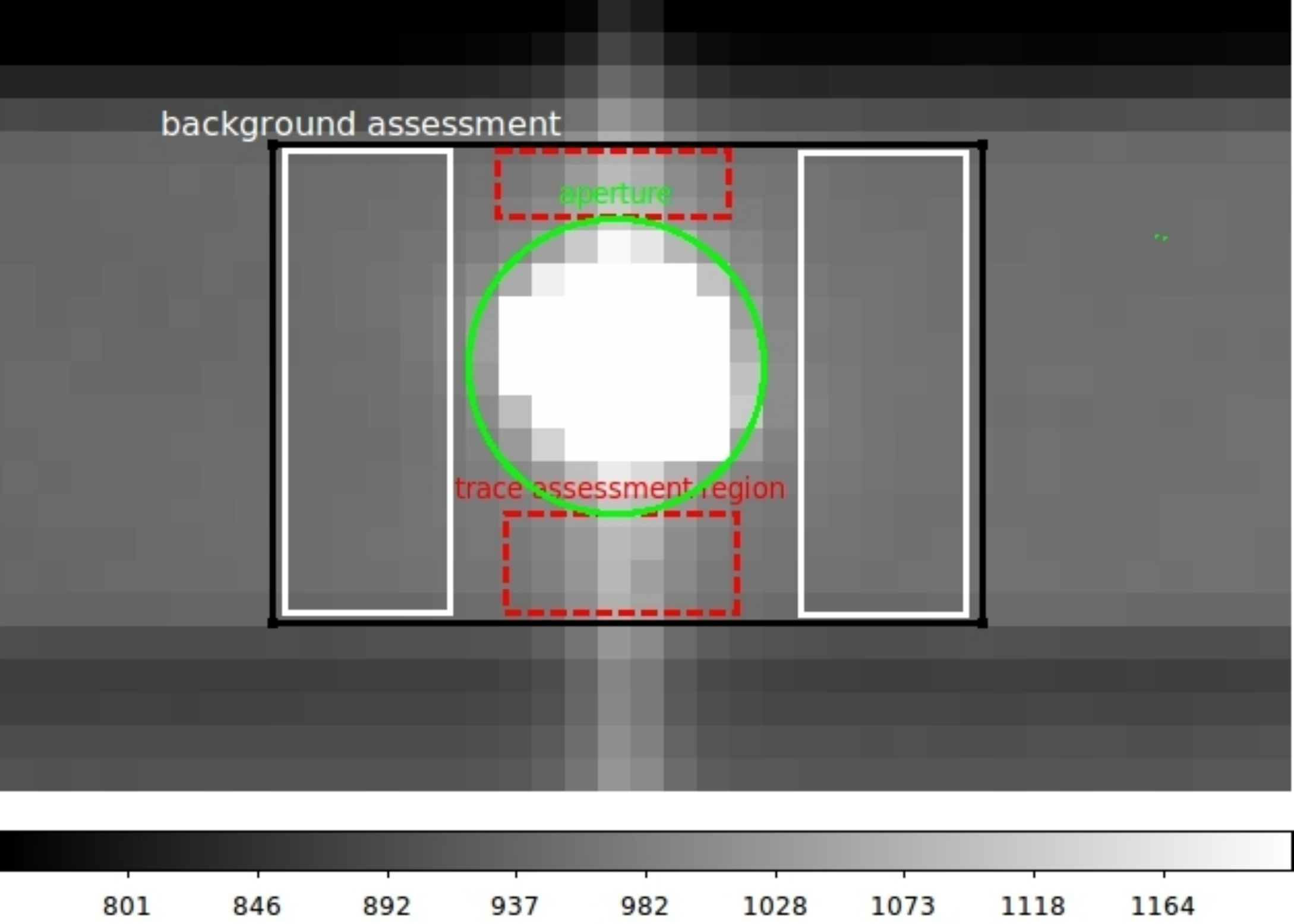}

\caption{Illustration of apertures, used for photometric measurement of stars. Circular aperture with radius 4.5 pixels were used to collect counts from a star. Rectangular regions to the left and to the right from the star was used to estimate the background level. Dashed rectangles were used to estimate the shape of the trail.}

\label{sel_areas}

\end{figure}

A typical images of the target star with a comparison star is shown in Fig.\ref{image}. Our reduction algorithm is similar to that of standard PySALT SLOTTOOLs package \citep{crawford10}, but with some differences. Photometric measurements of target and comparison stars were performed with fixed apertures around best fit centroids of the two stars, determined in each frame. 
We assume the background consists of a constant value and the vertical trail of the star, created by light from the star trailing during frame transfer from the image to the storage area of the CCD (the shutter remains open in Slot Mode). Typical selection regions are shown in Fig.\ref{sel_areas}. 
The background value was measured as a  median outside 3.75 arcsec aperture around stars and within 8.33 arcsec square, see Fig.\ref{sel_areas}.
Our measurements show that the vertical trail contains about 4\% of signal in 4.5 pixel star aperture, thus it should be carefully taken into account. We have adopted the following approach: We measured the mean shape of the trail profile over the areas, shown in Fig.\ref{sel_areas}. This profile was extracted from each frame row.

As a first step of correction for the atmospheric influence we have divided the flux of the target star by the flux of a comparison star 2MASS J12523138-2914023 (differential photometry).

\begin{figure}

\includegraphics[width=\columnwidth]{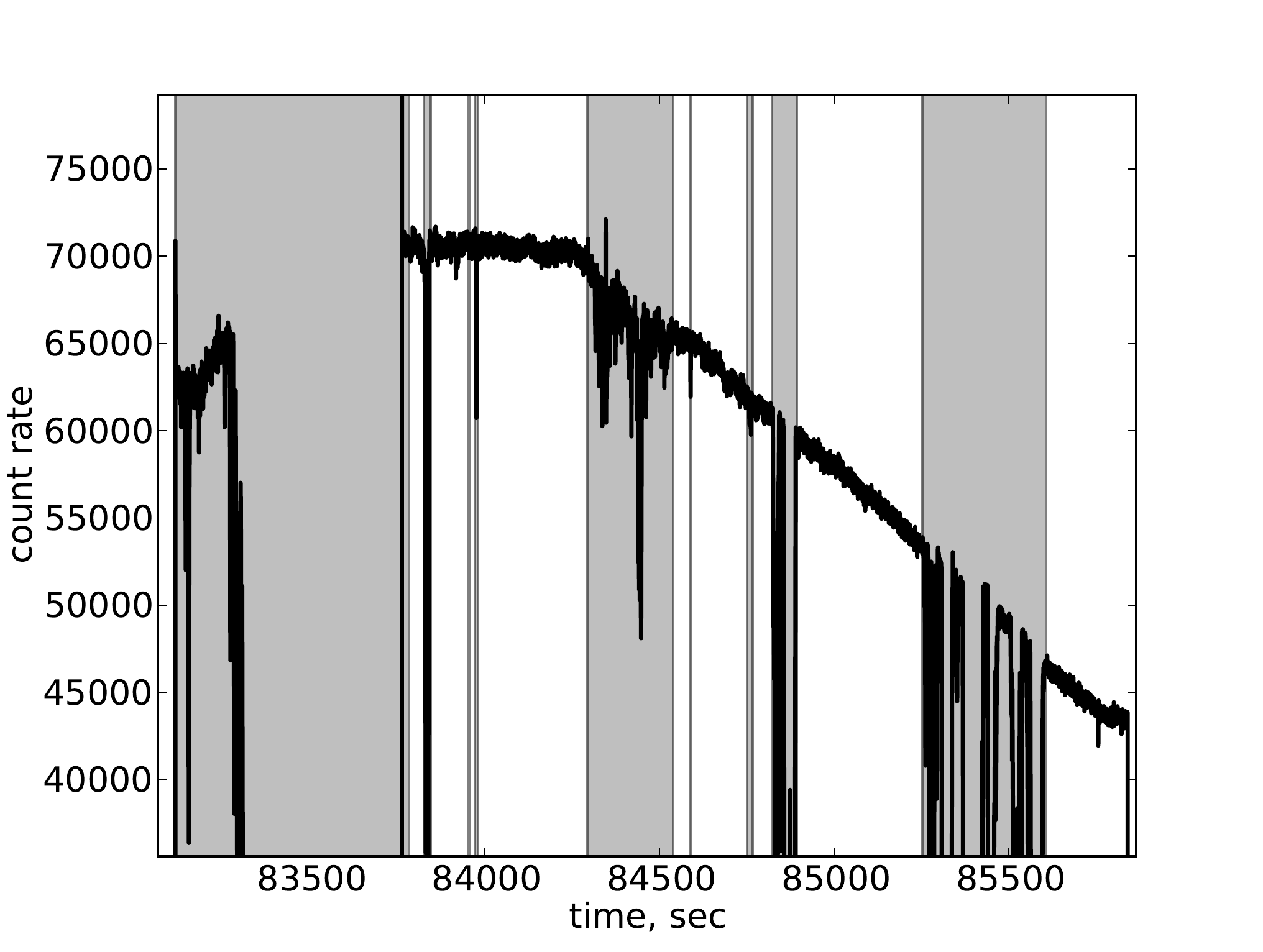}

\caption{Count rate of a comparison star during our observation. Due to poor weather condition some parts of our observations were affected by clouds. We have excluded time periods, shown by shaded areas from the subsequent analysis.}

\label{time_intervals}

\end{figure}
\begin{figure}

\includegraphics[width=\columnwidth]{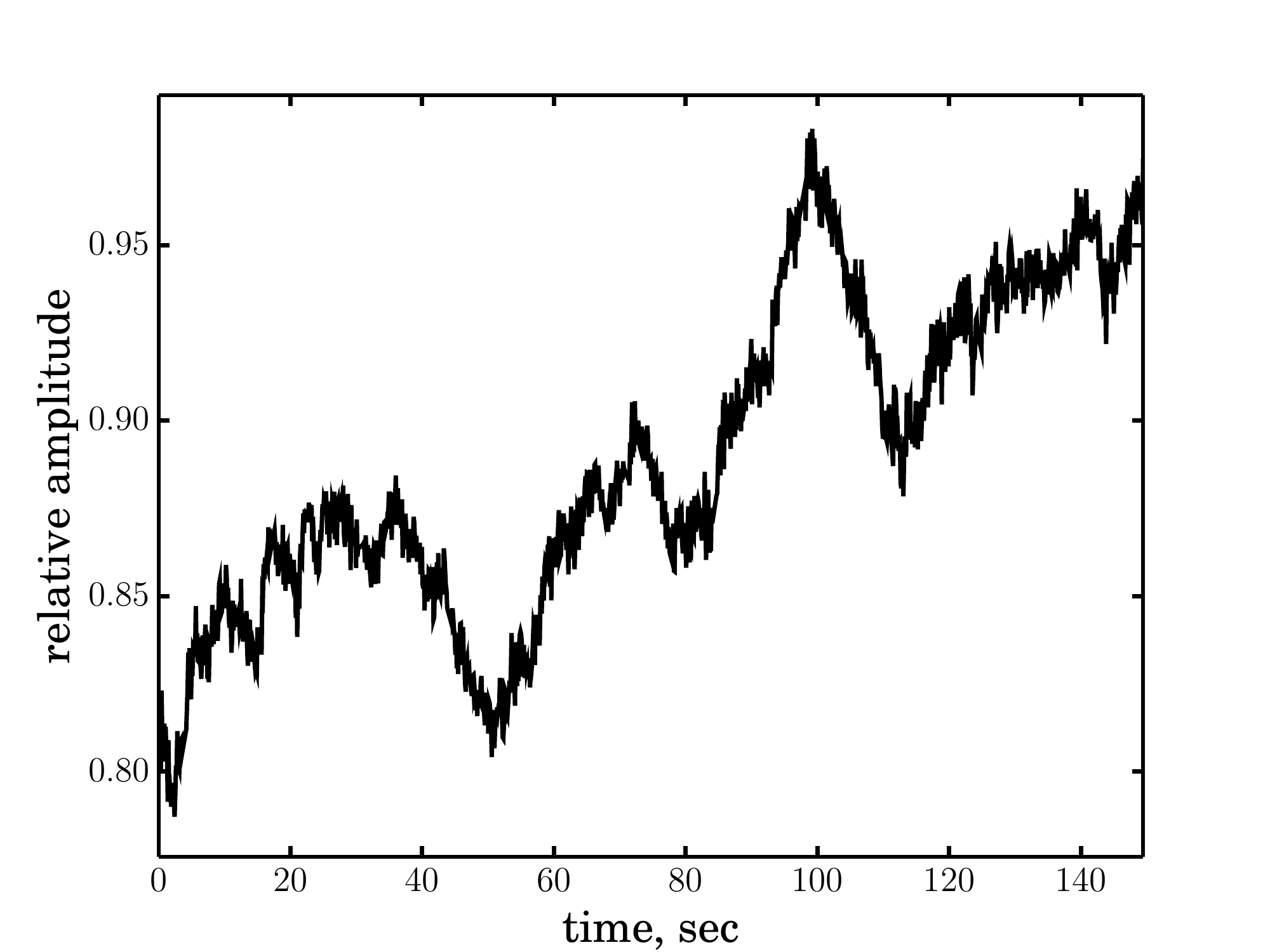}

\caption{A section of a light curve of EX Hya, which demonstrate aperiodic variations of the light down to at least 1 s timescale.}

\label{lcurve}

\end{figure}

Due to poor weather conditions, a substantial fraction (approximately 50$\%$) of our observations are not acceptable for the subsequent analysis. We have determined this from the light curve of the comparison star. The raw count rate of the comparison star is shown in Fig.\ref{time_intervals}, where excluded time intervals are shown by shaded areas. The remaining data were split into evenly sampled time intervals, each 6 s long. 
The exposure time of our dataset filtered for the bad weather conditions is 1340 seconds. Example of the source light curve is shown in Fig.\ref{lcurve}. It is clearly seen that aperiodic variability of the source flux exists down to at least 1 s timescale.

Numerous gaps in SALTICAM dataset (typically every 6 seconds, are due to features in the readout sequence) preclude usage of simple Fourier transforms to obtain power spectrum of the source variability. Thus we have obtained the power spectrum in two steps, at frequencies above 1/6 Hz and at frequencies below 1/13 Hz.

Seeing as the power spectrum at Fourier frequencies above 1/6 Hz has a steep slope of around -2 ($P(f)\propto f^{-2}$) \cite[see e.g.][]{revnivtsev11}, the simple rectangular window function can produce artificial power at all relevant frequencies. In order to reduce this effect, we have multiplied data in each time interval by a Hann window function before calculation of the power spectra. This multiplication was done for all datasets from all instruments.

\begin{figure}
\includegraphics[width=\columnwidth]{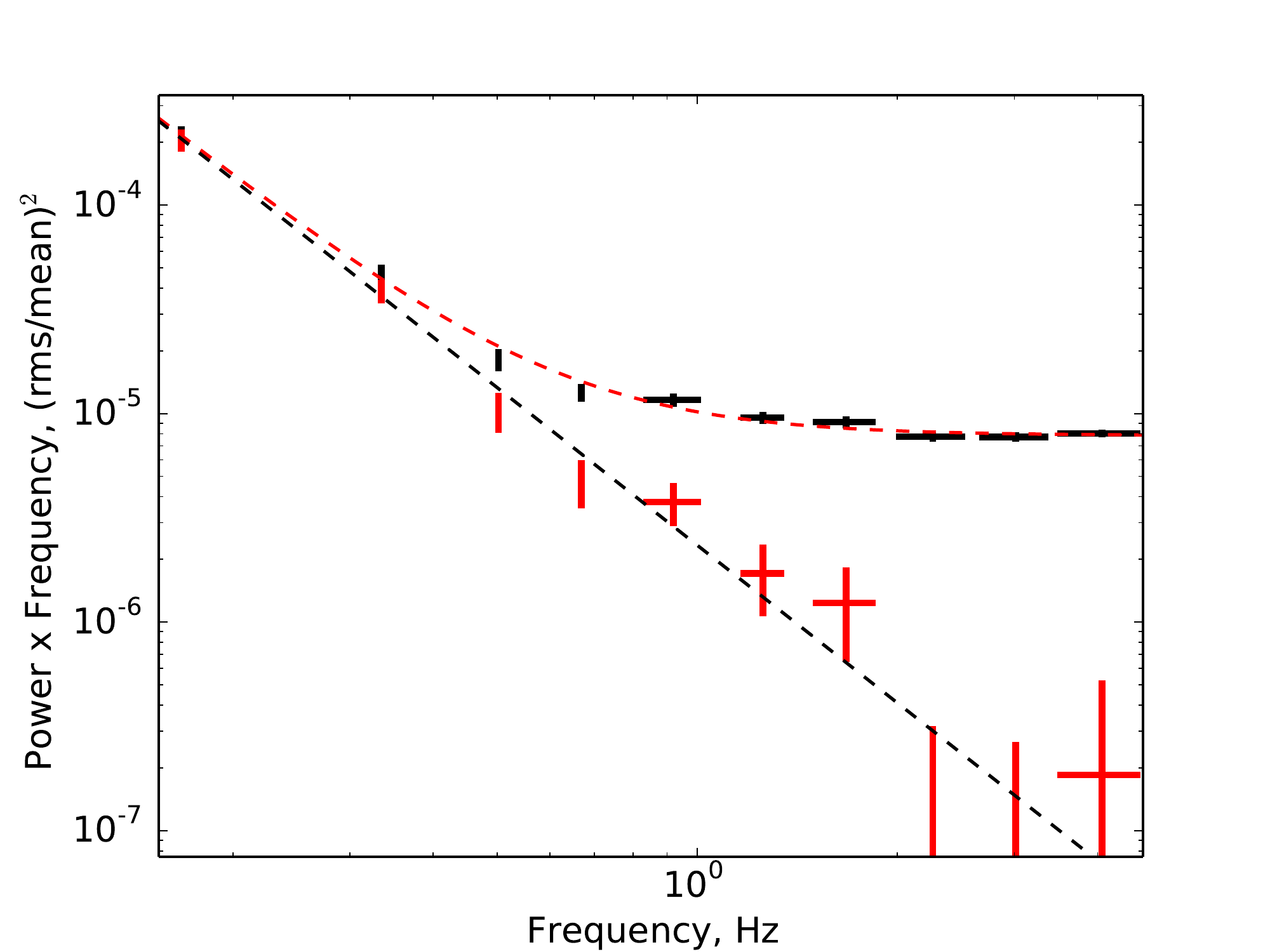}
\caption{Power spectrum of EX Hya, obtained by averaging over power spectra of all 6 s time intervals. The upper set of crosses show raw power density spectrum, which contains contribution from Poisson noise of photon counts. The lower set of crosses shows intrinsic power density spectrum of EX Hya. The dashed line shows a power law with slope $\alpha=2.5$}

\label{pds_salt}

\end{figure}

The resulting power density spectrum of EX Hya (in relative units (rms/mean)$^2$ Hz$^{-1}$, see e.g. \citealt{miyamoto91}) obtained by averaging over all power spectra, calculated for every time interval, is presented in Fig.\ref{pds_salt}. Uncertainty in the power value at any particular frequency interval was calculated from dispersion of all values of power, which fall into this interval. Such an approach provides us the most robust estimate of the uncertainty. This automatically includes all measurement errors and all uncertainties connected with intrinsic red noise in the source flux variations. This approach was also applied to all datasets, which we describe below.

The periodogram clearly contains the intrinsic source variability at frequencies below 1 Hz and is dominated by  noise at higher Fourier frequencies. The dominant contribution to the noise component is Poisson noise of photon counts, which we see as a constant value in units (rms/mean)$^2$ Hz$^{-1}$ (at the level approximately $\sim 8\times10^{-6}$ (rms/mean)$^2$ Hz$^{-1}$). 

Atmospheric turbulence along the light path adds a band limited noise (scintillation) to power spectra of optical objects \cite[see e.g.][]{dravins98,revnivtsev12}. It is known that the amplitude of this noise is smaller for larger telescopes \citep{dravins98} and in the case of SALT/SALTICAM observations we do not detect it. 
In order to estimate an upper limit to the atmospheric scintillation component we fit the power spectrum using model with additional component $P_{\rm atm}=A(1 + (f/f_0)^2)^{-0.3}$, $f_0 \sim 0.8 \rm Hz$ .
The upper limit on contribution of atmospheric scintillations at 1 Hz is about $\sim5\times 10^{-6}$ (rms/mean)$^2$ Hz$^{-1}$.

\subsection{SALT/BVIT}
	BVIT - The Berkeley Visible Image Tube is a photon counter with quantum efficiency about $\sim 15 \%$ and angular resolution approximately $0.3''$, mounted on SALT \citep{buckley10,welsh12}. BVIT makes observations in the format of 25 nsec time resolution event list in a circular, 1.6-arcmin diameter field of view.  The target, the background and the comparison star apertures were selected during data processing. We have binned all count rates into  0.01 and 0.1 sec time resolution bins. All BVIT data processing made use of the  BVIT data standard processing tools \citep{welsh12} 
	
The mean BVIT count rate of EX Hya  with background is 5.9 kcps, while the mean background count rate is 0.9 kcps. The background rate is not constant and contains shot noise variations as well as long term trends. Point-to-point subtraction of the background curve from the count rates of EX Hya will increase the shot noise of the resulting light curve, therefore it is better to subtract a model of the background count rate variations. Analysis of the time variable background count rate shows that a third order polynomial fit describes it well.
	
\begin{figure}
\includegraphics[scale=0.4]{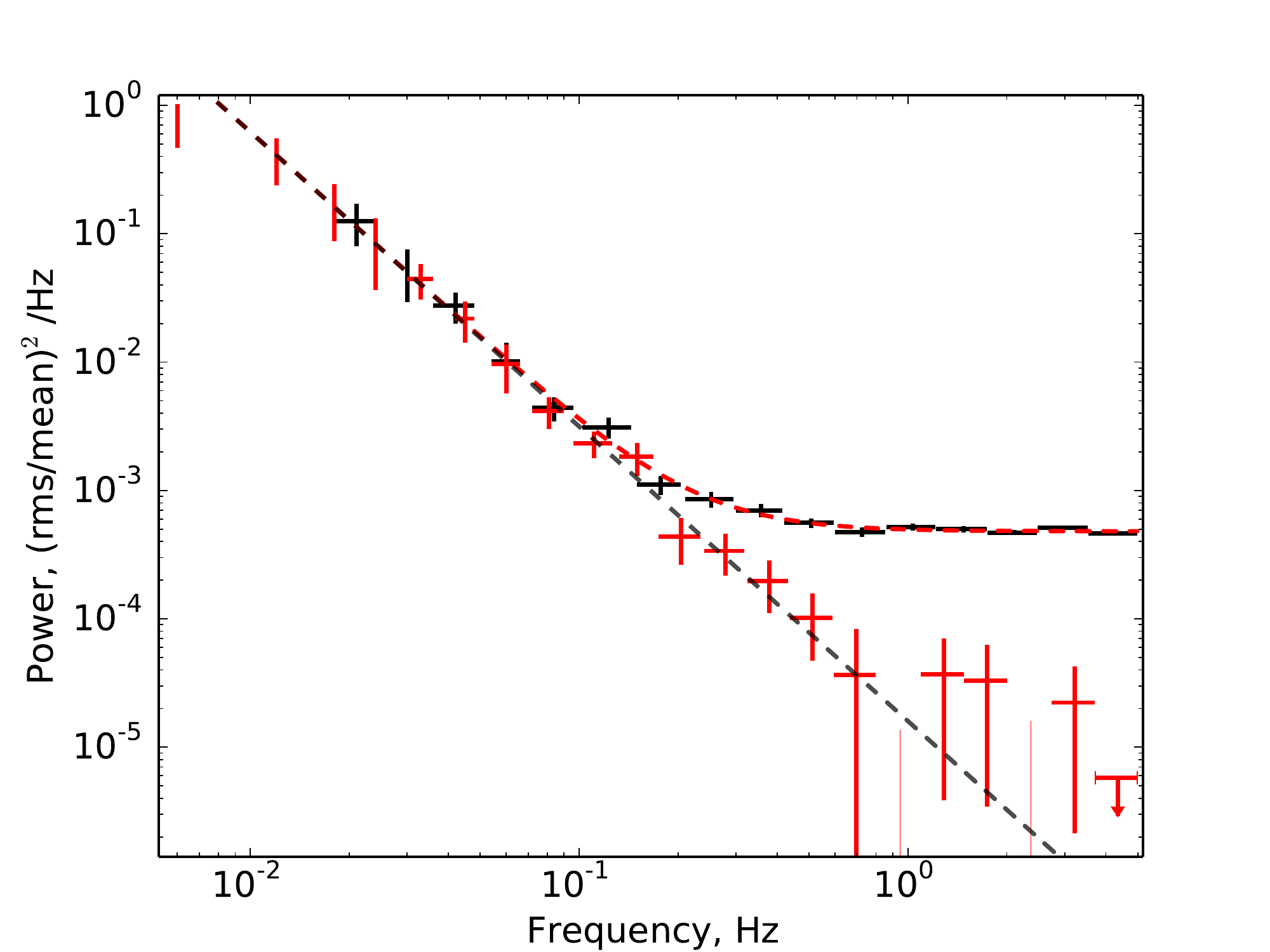}
\caption{Power spectrum of EX Hya light curve received by SALT/BVIT. The dashed line shows a power law with slope $\alpha=2.48$}
\label{BVIT}
\end{figure}

The resulting power spectra of EX Hya from BVIT data (original and shot-noise-subtracted) are presented in Fig.\ref{BVIT}. 
		
\subsection{1.9m/SHOC} 

SHOC (Sutherland High-speed Optical Camera) consists of an Andor iXon 888 CCD camera (similar to that used by our group on RTT150 telescope \citealt{revnivtsev12}) and auxiliary systems \citep{coppejans13}. For our observations SHOC was mounted on the SAAO 1.9 m telescope. 
SHOC was used in both conventional and electron-multiplying modes, with binning optimized to the seeing conditions on each night.

Instrumental magnitudes for EX Hya were extracted using IRAF DAOPHOT \citep{stetson87} tasks for different apertures. Aperture-corrected photometry was then performed by using the IRAF MKAPFILE task \citep{davis93}, extracting the magnitudes for the smallest apertures that maximize the signal-to-noise and ignoring data points with errors larger than 0.1 mag. The data were then filtered to eliminate outliers, which were detected using a simple algorithm which marks as an outlier any point outside of 5 sigma from the median of 5 neighbouring points. The sigma parameter for this filter was the standard deviation of 100 points preceding the given one, which were initially detrended by making a linear fit.

\begin{figure}
\includegraphics[scale=0.4]{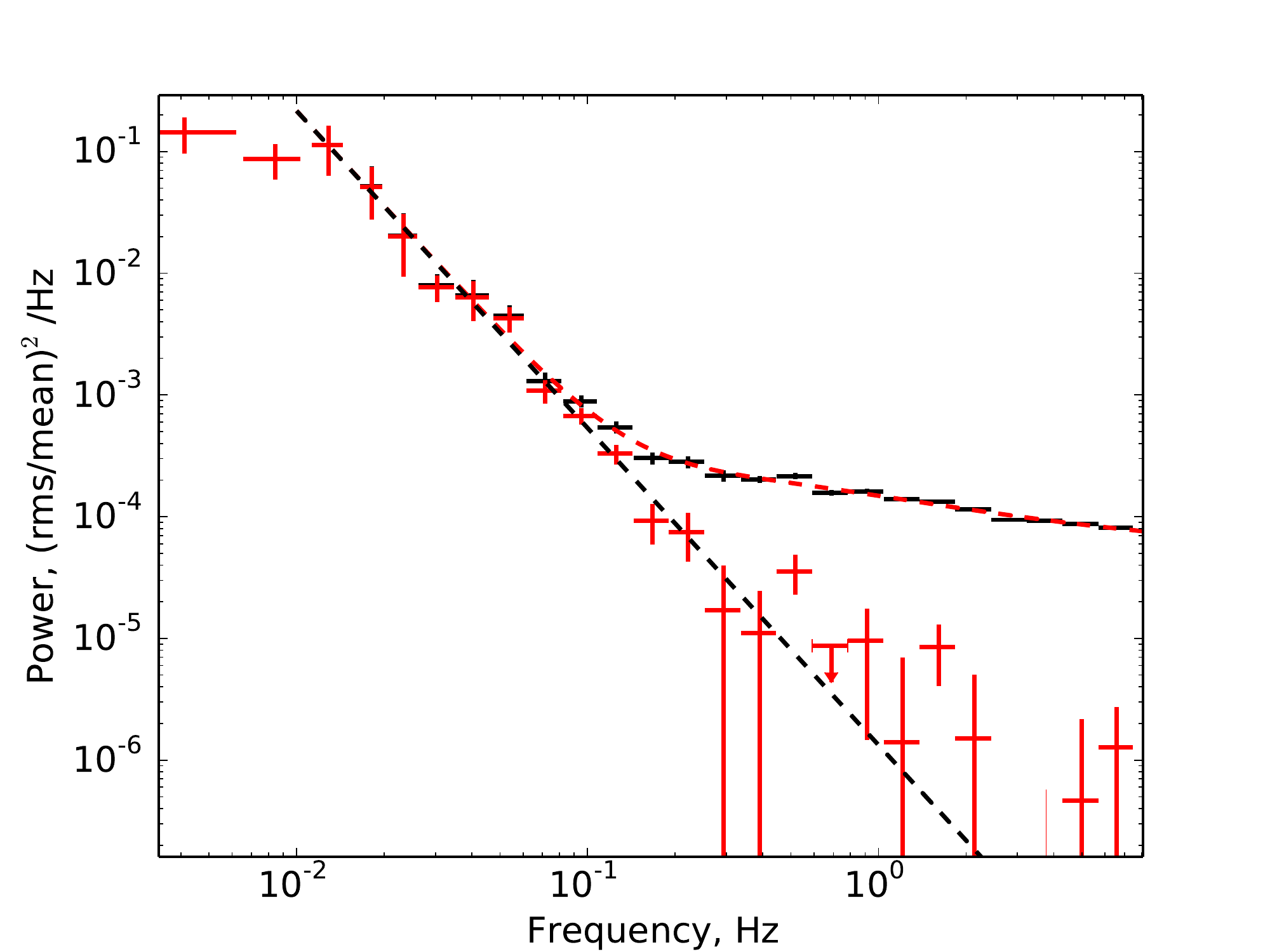}
\caption{Power spectra of EX Hya light curve, obtained in one 1.9m/SHOC observations. Upper set of points shows raw power spectrum, which contains photon counting noise and noise of atmospheric scintillations. Lower set of points shows EX Hya power spectrum with subtracted noise components. The dashed line shows a power law with slope $\alpha=2.48$}
\label{SHOC_atmosphere}
\end{figure}

The power spectrum of EX Hya, obtained by SHOC, contains noise from atmospheric scintillations which has to be taken into account. Power of noise of atmospheric scintillations $P_{\rm atm}(f)$ as a function of Fourier frequency $f$ was modelled by a simple analytic function of the form:
\begin{equation}
 P_{\rm atm}(f) \propto (1 + (f/f_0)^2)^{-0.3} 
\end{equation}
An example of a power spectrum of EX Hya (raw and noise subtracted), obtained in one of the SHOC observations, is shown in Fig.\ref{SHOC_atmosphere}.

\subsection{1.9m/HIPPO}

For completeness we also used light curves of EX Hya obtained on the SAAO 1.9m telescope using the HIPPO photo-polarimeter \citep{potter08}, presented in work of \cite{revnivtsev11}.
HIPPO is a 2 channel instrument capable of simultaneous dual filtered photo-polarimetry. We present results of light curves collected in one filter (R). Data reduction was done as outlined in \cite{potter10} and binned to 1 second time resolution.

\section{Results}

\begin{figure}
\includegraphics[width=\columnwidth]{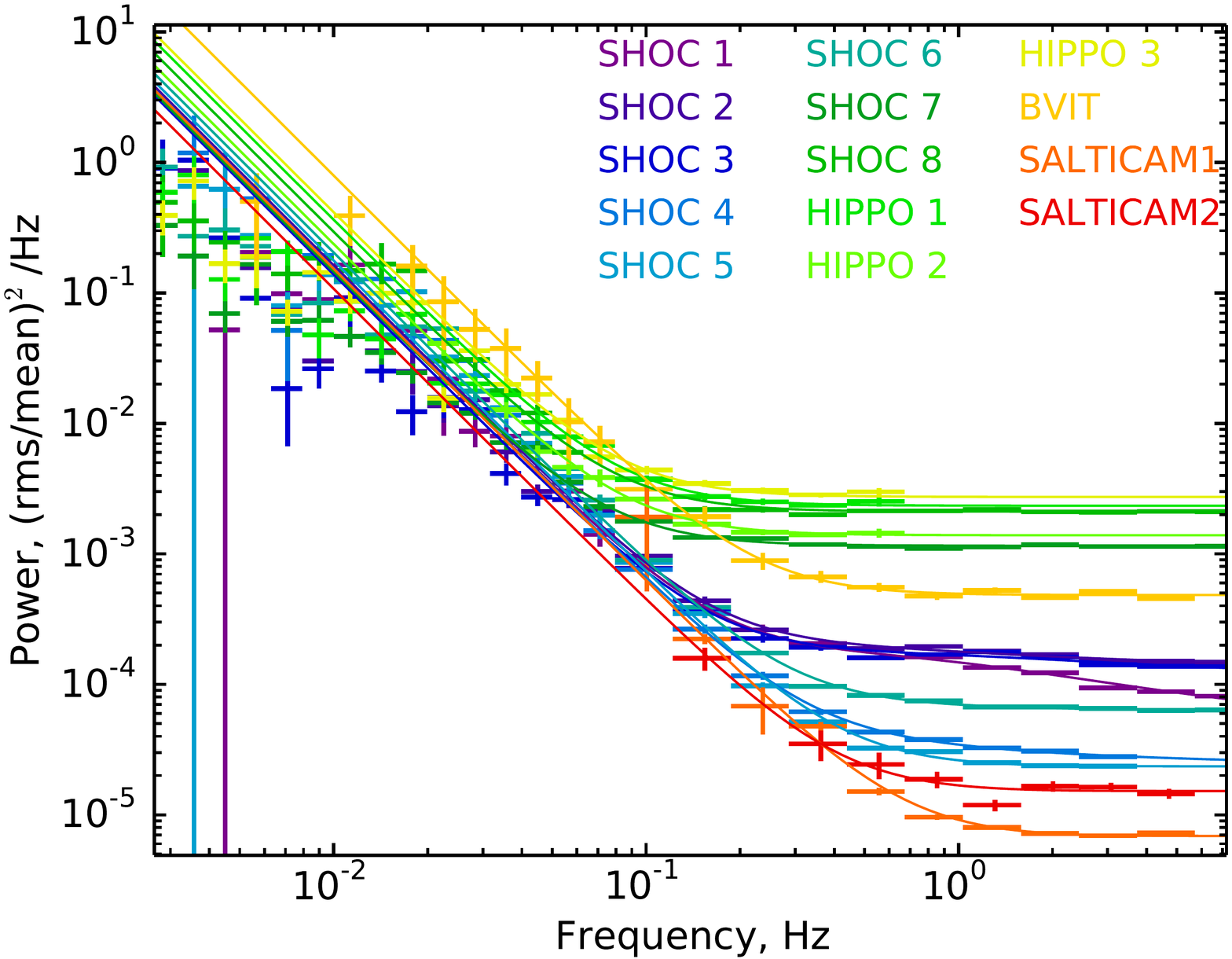}
\caption{Power spectra of EX Hya, obtained during different observations with different instruments along with their best fit models (solid curves)}
\label{total_pds}
\end{figure}

Power spectra from observations obtained at the SAAO with SALT/SALTICAM, SALT/BVIT, SAAO 1.9m/SHOC and SAAO 1.9m/HIPPO (shown in Fig.\ref{total_pds}) were analysed at frequencies above 0.05 Hz. All SHOC observations were split into 8 parts which have approximately similar levels of atmospheric oscillations and Poisson noise. SALTICAM data were split into two parts with similar levels of Poisson noise.

Power spectra were fitted using a method of maximum likelihood on unbinned values of power \cite[see e.g][for its application to power spectra]{barret12}. In short, we have maximized the likelihood $L$, which is:
$$
L=\prod_{ij} {1\over{S_{ij}}}\exp(-P_{ij}/S_{ij})
$$
where $S_{ij}$ is the model of the source variability power at frequency $f_{j}$ for lightcurve segment $i$, $P_{ij}$ is the measured value of the power at frequency $f_j$ from analysis of lightcurve segment $i$. The model $S_{ij}$ is a function of Fourier frequency $f_{j}$, the break frequency $f_{\rm break}$, the slope of the power law before the break $\alpha$, normalization constants $C_i$ (they might be different for every separate lightcurve segment $i$), and noise $P_{\rm atm}+P_{\rm Poiss}$ (atmospheric scintillations and Poisson noise).

$$
S_{ij}= C_{i} f_j^{-\alpha}[1 + (f_j/f_{\rm break})^2]^{-2}+P_{\rm atm}+P_{\rm Poiss}
$$

Atmospheric scintillation noise in the form $P_{\rm atm}=A_i(1 + (f_i/f_0)^2)^{-0.3}$ was found to be statistically significant only for 1.9m/SHOC data.

\begin{figure}
\vbox{
\includegraphics[width=\columnwidth]{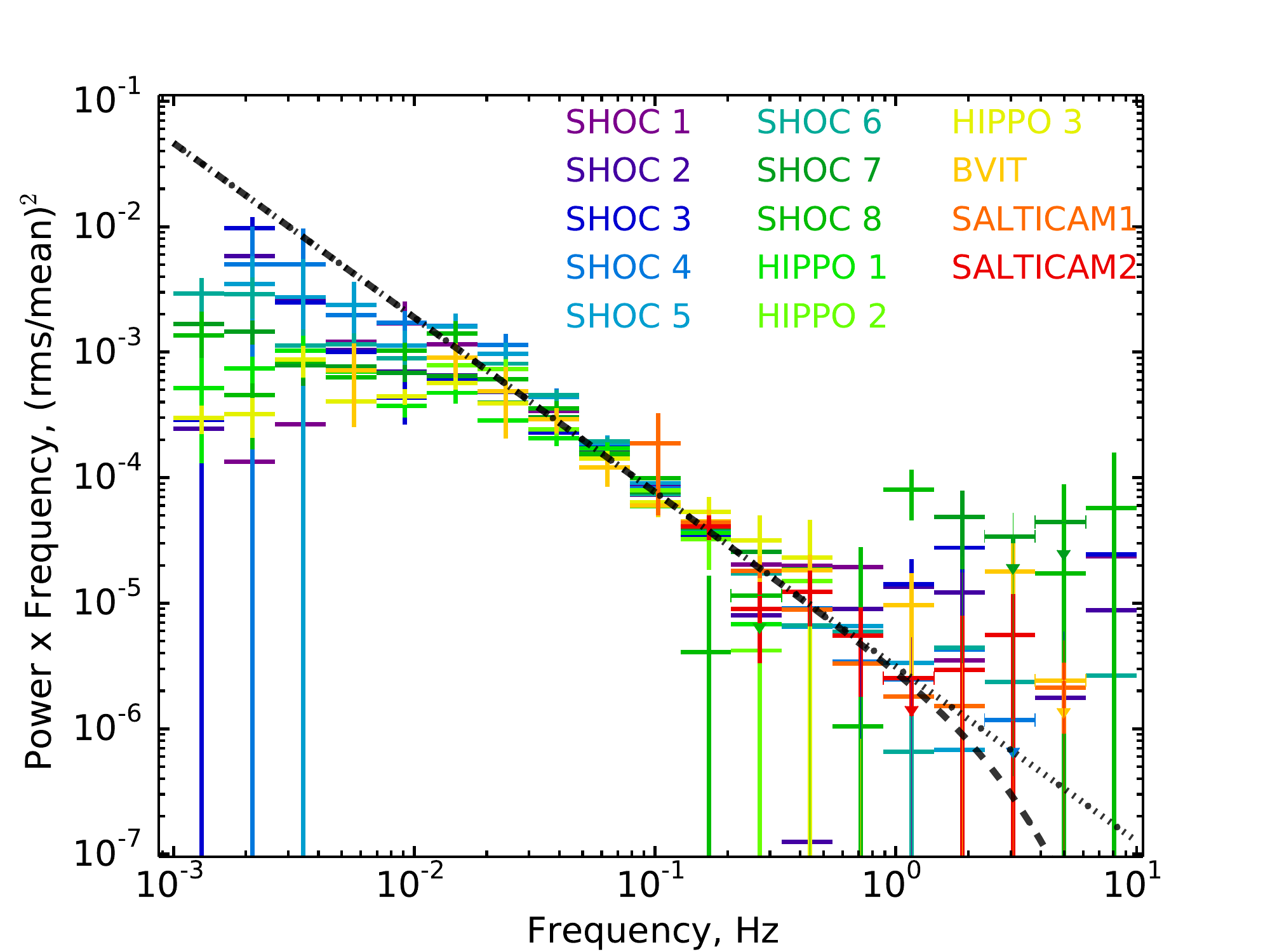}
\includegraphics[width=\columnwidth]{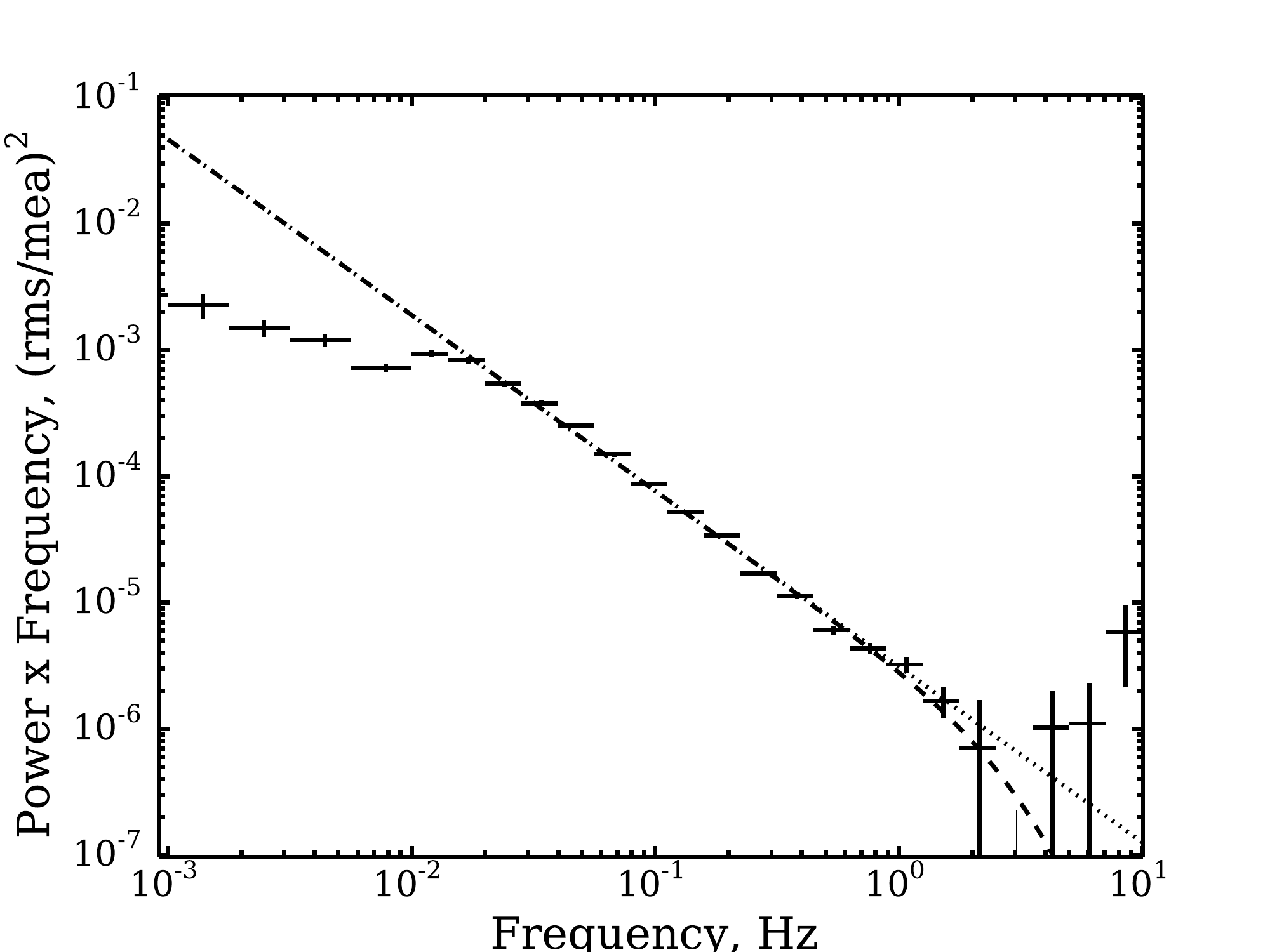}
}
\caption{Broad band power density spectrum of aperiodic variations of optical light of EX Hya, obtained by different instruments (upper panel) and averaged over all presented measurements (lower panel). Dotted line shows a power law with the slope $\alpha=2.4$. Dashed curve (with rollover at high frequencies) shows the same power law with a break at $f_{\rm break}=4.4$ Hz.}
\label{pds_broad}
\end{figure}

Power spectra of all separate lightcurve segments, renormalized to the average normalization, are shown in Fig. \ref{pds_broad}. For this plot we have subtracted fitted contributions of all noise components.

In order to present the frequency binned power spectrum of EX Hya averaged over all analysed datasets, we also used maximum likelihood technique. In this case the mean value of the power in any individual frequency bin was estimated by maximizing the likelihood function assuming that the model function is constant within this frequency bin. The confidence interval on the mean value of the power in this frequency bin was obtained adopting intervals of power values where $\Delta \log L=0.5$.

The power spectrum, obtained after averaging over all available datasets is presented in Fig.\ref{pds_broad}. The power spectrum does not show any evidence of quasi-periodic oscillations (QPOs), with a typical upper limits of 0.3\% . The best fit approximation of obtained datapoints give the value of the slope $\alpha=-2.40\pm0.05$, the 2$\sigma$ lower limit on the break frequency $f_{\rm break}>3.5$ Hz.
The quality of the fit can be accessed via calculation of $\chi^2$ values on binned power spectra. For the PDS averaged over all dataset the $\chi^2=13.5$ for 12 degrees of freedom.

\subsection{Area of accretion curtains footprints}

The post-shock plasma cooling timescale allows us to estimate its density and thus the area of the accretion curtain footprints. The simplest estimate of the cooling time \citep{langer81}:

\begin{equation}
  \tau \sim {E_{\rm ps}}/{Q_{\rm ps}}
\end{equation}

where $E_{\rm ps} = {n_{\rm ps} k_b T}/{(\gamma - 1)}$ is thermal energy density, $Q_{\rm ps} \sim k_{\rm bremss} n_{\rm ps}^2 \sqrt{T_{\rm ps}}$  is unit volume cooling rate of plasma in the post-shock region with density $n_{\rm ps} = \rho_{\rm ps}/m_{\rm p} \approx 4 {\dot{M}}/({v_{\rm ff}A m_{\rm p}})$ and temperature $T_{\rm ps} \approx {3 v_{\rm ff}^2 m_{\rm p}}/({16 k_b})$. Here $k_{\rm bremss} = 1.4 \times 10^{-27}$ is the bremsstrahlung cooling rate coefficient \citep{rybicki}, $A$ is the cross-sectional area of the accretion curtain, and $v_{\rm ff}$ the free fall velocity of the matter just above the shock front. It is easy to see that an upper limit on the cooling time provides an upper limit on the accretion curtain cross sectional area $A$.  
\begin{equation}
 A \approx \frac{8 \tau (\gamma - 1) k_{\rm bremss} R_{\rm WD} \dot{M}}{\sqrt{3 k m_{\rm p}^3} G M_{\rm WD}} 
\end{equation}
\begin{equation}
 A \approx 1.3 \cdot 10^{15} \tau \frac{R_{\rm WD}}{10^9 \textrm{cm}} \frac{M_\odot}{M_{\rm WD}}\frac{\dot{M}}{10^{15} \textrm{g s}^{-1}} ~~\textrm{cm}^2
\end{equation}
Here $M_{\rm WD}, R_{\rm WD}$ are the white dwarf mass and radius respectively, $k_{\rm bremss}$ the bremsstrahlung cooling coefficient, $\gamma$ the adiabatic parameter ($\gamma = 5/3$ was used for our computation), $\dot{M}$ the mass accretion rate, $\tau$ the hot zone cooling time.  

From this equation, the area of both curtains (accretion is going onto two poles) is $A<10^{15}$ cm$^2$ and the specific mass flux is about $\dot{M}/A>3$ g sec$^{-1}$ cm$^{-2}$.
This simple approach underestimates the cross-sectional area due to the underestimating of the cooling rate in the hot zone. A more accurate estimate of the cross-sectional area $A$ can be derived by taking into account the profiles of the temperature, velocity and density in static hydrodynamic flow through the accretion column \citep{wu94}. The effect of gravity in the column and geometrical compression, which can play significant role in the case of tall accretion columns, also should be taken into account \citep{Canalle05}. 

We use the \cite{Canalle05} one dimensional equations of the stationary flow (eq. 37, 38 in the paper). We assume that accreting matter flows from the inner disc radius along the magnetic field lines through the geometrically thin curtains to the two areas on the opposite sides of the white dwarf. Each curtain collects half of the accreted matter. Adopting the system parameters from Table \ref{pars}, the minimum possible cooling mechanism ($\Lambda_{\rm bremss} \sim 1.4 \times 10^{-27} n^2 \sqrt{T}$) and cooling time ($\tau< 0.3$ sec) we get for a single curtain footprint a surface area  $A < 10^{15}$ cm$^2$ and a mass accretion rate of $\dot{M}/A> 3$ g s$^{-1}$ cm$^{-2}$. 

 The value of the specific mass accretion rate, which we estimate here, is very important. Usually this parameter is an ingredient of all models of X-ray spectra of IPs \cite[see e.g.][]{cropper98,suleimanov05,yuasa10}. However, it is highly degenerate for observational data with limited energy resolution (i.e. if there are no abilities to make detailed emission line diagnostics). This degeneracy usually leads to adopting some fiducial parameters of the specific mass accretion rates. Here we are able to put physically important constraints on this parameter.

It should be noted that with the obtained parameters the accretion curtains might become optically thick with respect to Compton scattering in the vertical direction (i.e. along the direction of mass flow). This might have an influence on the spectral shape of the emission of the post-shock region \cite[see e.g.][]{suleimanov08} and on its angular dependence, producing pulsed emission.

The length of the accretion curtain can be estimated from eclipse mapping. In particular, \cite{mukai98} showed that the  egress and ingress of eclipses of X-ray emitting regions last about 21 sec, which can be recalculated into a maximal spatial size of the X-ray emitting accretion curtain near the WD surface of $l\sim 10^9$ cm. For this length, the accretion curtain thickness should then be $\la  10^6$ cm.

\begin{figure}
\includegraphics[width=\columnwidth,bb=29 21 528 396,clip]{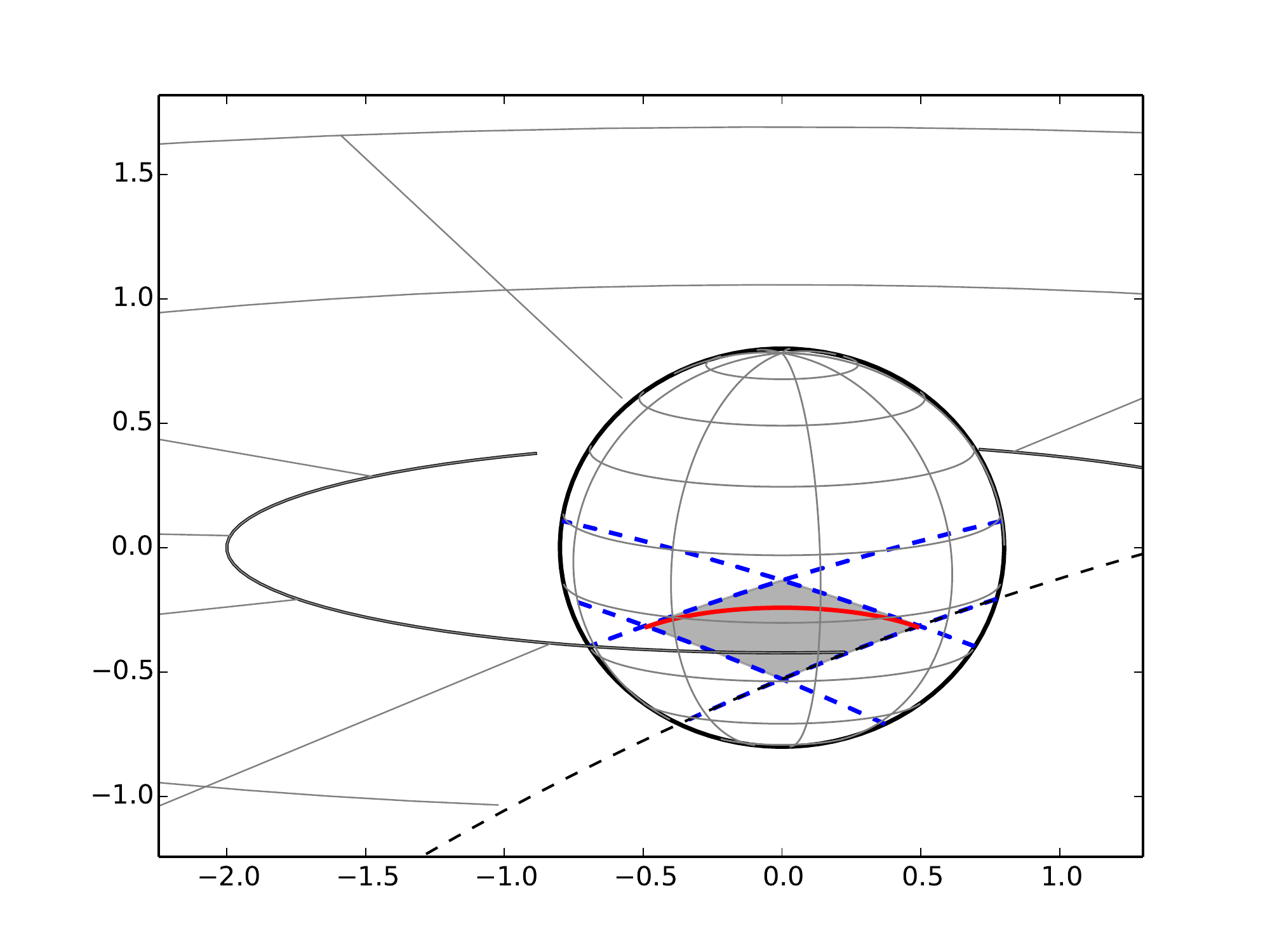}
\caption{Schematic view of the accretion curtain geometry. The scheme illustrates that the area of the rhombus, which limits the size of the accretion curtain footprint and which is determined from the eclipse mapping, can be significantly larger than the real area of the accretion curtain -- thin stripe within the rhombus. Part of big dashed circle (at the lower right corner of the plot) denotes the secondary at the position immediately after the egress of X-ray eclipse.}
\label{eclipse_scheme}
\end{figure}

It is important to emphasize here that the area of the accretion curtains footprints, which we measure here with the cooling time method, is significantly smaller than that, estimated from eclipse mapping \cite[see e.g.][]{hellier97,odonoghue06}. However, there are no contradictions here because the eclipse mapping essentially determines the maximal spatial size of the accretion curtain and not its area (in a sense it determines the upper limit on the area of the column). As an illustration of this difference, in Fig.\ref{eclipse_scheme} we show a schematic of the accretion curtain footprint along with limits of its sizes from the eclipse mapping.

The deduced parameters of the accretion curtains imply that the plasma density in the curtains increase from $n\sim10^{16}$ cm$^{-3}$ to $n\sim10^{17}$ cm$^{-3}$ in its luminous part. We postulate that this range of hot plasma ($T\sim(50-200)\times 10^6$ K) densities can be probed with iron emission lines diagnostics \cite[see e.g.][]{hayashi13} if high energy resolution spectra of the curtains will be available from future calorimeter experiments  aboard e.g. satellites Astro-H \citep{takahashi10} or Athena+ \citep{denherder12}.

\subsection{Matter coupling region at the magnetospheric boundary}

Assuming that the matter flows strictly along the magnetic field lines of the WD dipole, we can relate the thickness of the accretion curtain at the WD surface with $\Delta r$, the radial extent of the plasma-magnetosphere coupling region at the inner edge of the accretion disc \cite[see e.g.][]{rosen92}. Assuming that the magnetic dipole and the accretion disc rotation axis are co-aligned (which is a reasonable assumption at the current level of accuracy of our estimates), the cross sectional area of one curtain corresponds to a depth $\Delta r/R_{\rm in} < 10^{-3}$. {\sl Note, that the size of this coupling region is significantly smaller than that assumed in many works on magnetic accretion} \cite[see e.g.][]{basko76,ghosh78,hameury86,buckley89,rosen92,kim95,ferrario96}.We can try to compare this number with some physical estimates of the plasma layer thickness on top of the magnetosphere \cite[see also discussion of this topic in][]{ichimaru78,ghosh78,ghosh79,spruit90,campbell92,lovelace95,campbell10}. 

It is commonly accepted that matter at the inner part of the accretion disc starts to flow along magnetic field lines when the (central object) magnetic field energy density exceeds the matter pressure. The flow along the magnetic field lines very rapidly becomes supersonic, but the motion of plasma across the field lines is strongly suppressed and thus is significantly subsonic in this direction.

Inevitably Bohm diffusion occurs of highly conducting plasma across the magnetic field lines (which is higher that classical cross-field diffusion, see e.g. review in \citealt{kadomtsev83}), with a diffusion coefficient 
\begin{equation}
 D_{\rm Bohm}\sim {1\over{16}} {kT\over{eB}}
\end{equation}

This will lead to thickening of any infinitely thin layer of plasma sheet to a length scale $\Delta r_{\rm Bohm}\sim \sqrt{D_{\rm Bohm} \tau}$, where $\tau \sim \omega_{\rm K}$ is the timescale of the plasma free-fall time from the inner disc to the WD surface ($\sim7-8$ sec, in the case of EX Hya, see \citealt{revnivtsev11}). This thickening can be related to the speed of sound in the accretion disc plasma $c_{\rm s}$: 
\begin{equation}
 \Delta r_{\rm Bohm}\sim {c_{\rm s}\over{\omega_{\rm K}}} \left({\omega_{\rm K}\over{\omega_{\rm c}}}\right)^{1/2},
\end{equation}
where $\omega_{\rm c}=eB/m_{\rm p}$  is the proton gyro-frequency. It is clear that in our case, with a relatively strong magnetic field $B\sim10^{5-6}$ G, this diffusive thickening of the plasma sheet is negligible in comparison with the thermal pressure scale of plasma in the accretion disc with $h\sim c_{\rm s}/\omega_{\rm K}$.

Thus we propose that the thickness of the plasma flow in the magnetosphere is set by its dynamical settling at the magnetospheric boundary. Different works, adopting reasonable assumptions about the accretion disc magnetic diffusivity, give estimates of $\Delta r \la c_{\rm s}/\omega_{\rm K}$ \cite[see e.g.][]{campbell92,shu94,lovelace95,campbell10}, which is in general agreement with results of direct numerical simulations \cite[e.g.][]{long05,romanova13}.

In the case of EX Hya the size of the magnetosphere ($\sim$3-5 $R_{\rm WD}$, \citealt{siegel89,hellier97,revnivtsev11}) is smaller than the co-rotation radius ($\sim50 R_{\rm WD}$). This means that there should be a strong shear ($v_{\rm m}/v_{\rm K}\sim 0.01$) between the magnetospheric boundary and the accretion disc plasma. This shear should significantly suppress the interchange (Rayleigh-Taylor) instability, often assumed to be operating in the case of accretion on slow magnetic rotators \cite[see e.g.][]{arons76,romanova08}.  Part of the excess rotational energy (which is not transferred to the rotational energy of the white dwarf) should result in some additional heating of the innermost part of the accretion disc. This additional energy release can not heat the matter to X-ray temperatures, because eclipse mapping of EX Hya showed that the X-ray emitting region in this system is confined to the WD surface \cite[e.g.][]{mukai98}. Thus, the effective temperature at the innermost edge of the 
disc should be less than  
$<$ 2 -- 3 eV ($<20-30\times 10^3$ K)\cite[][]{mauche99}.

Adopting the other parameters of our binary we can conclude that the thickness of the inner disc should be $H/R \la 5\times10^{-3}-10^{-2}$. Thus the size of the coupling region at the boundary of the magnetosphere is $\Delta R/R\la H/R\la 5\times10^{-3}$, which is in agreement with our findings.

\subsection{Impact for neutron stars}

Accretion flow onto magnetic neutron stars should form magnetosphere-disk interaction regions quite similar to those of accreting magnetic white dwarfs. The size of the WD magnetosphere in EX Hya is $\sim1.9\times 10^{9}$ cm. This is close to size of magnetosphere of well known magnetic neutron star accretor EXO 2030+375 in its faint state \cite[e.g.][]{reig98}. An estimate of the magnetic moment $\mu$ of the WD in EX Hya $\sim 2.5 \times 10^{30}$ G cm$^{-3}$ is close to that of neutron star in EXO 2030+375 $\sim (2-10)\times 10^{30}$ G cm$^{-3}$ \cite[e.g.][]{klochkov07}, thus the magnetic field strength at the boundary of the accretion disk in these systems are of the same order. The mass accretion rate in faint state of EXO 2030+375 (with $L_{\rm x}\sim 10^{36}$ erg sec$^{-1}$) is only few times higher than that in EX Hya. The only difference is the higher level of illuminating X-ray flux from accreting neutron star in EXO 2030+375, which might lead to some additional heating of the innermost parts of the accretion disk.

Therefore, we can anticipate that the overall structure of the compact object magnetosphere -- accretion disk interaction in these two binaries should be similar and thus, the thickness of the accretion flow plasma/magnetosphere coupling region should be of the same order $\Delta R/R < \textrm{few}\times 10^{-3}$.
In this case the area $A$ of the accretion column/curtain on the neutron star surface should be quite small. Assuming that one accretion curtain footprint occupies sector $\pi$ of annulus of the radius $R_{\rm NS}\sqrt{R_{\rm NS}/R_{\rm m}}$ over the neutron star surface around its magnetic axis, its fractional area can be estimated as follows:

\begin{equation}
 {A\over{4\pi R_{\rm NS}^2}}< {1\over{4}} {R_{\rm NS}\over{R_{\rm m}}} {\Delta r\over{R_{\rm m}}}\sim 10^{-6}
\end{equation}

Even in the case of not luminous accretion powered neutron stars, such a small fractional area would result in very high mass flux, and thus in accretion flow which is (locally) highly super-Eddington. This parameter might have a strong influence on the structure of the accretion column in such neutron stars and thus merits further exploration.

\section{Summary}

In this work we are trying to determine the area of accretion curtain footprints on the surface of the accreting magnetic white dwarf in the binary system EX Hya. Our approach is based on measurements of the hot post-shock plasma cooling time.  The plasma in the accretion curtains, which is heated in the standing shock above the WD surface, cools on a timescale which is determined by its density. All variations in the mass accretion rate at the WD surface at timescales smaller than this (and we know that such fast mass flux variations do exist in accreting neutron star binaries) should be effectively smeared out in the luminosity variations. Knowing the values of the total mass accretion rate at the WD surface, the hot plasma temperature and cooling time, allow us to make an estimate of the accretion curtain footprint area.

We have attempted to measure the smearing timescale with the help of fast timing observations of one of the brightest intermediate polars, EX Hya, with optical telescopes at the South African Astronomical Observatory. The principle is based on the fact that a significant part of optical emission from EX Hya originates as reprocessed X-rays in the post-shock region of the WD accretion curtains. 

We have used a set of 9 observations on different telescopes and produced a power spectrum of EX Hya in Fourier frequency interval $\sim 10^{-3}-5$ Hz. Unfortunately the total exposure time of our best quality highest time resolution dataset, from SALT,  is only 1340 seconds. However, the quality of the SALT data shows that we can obtain good quality power spectra at high frequencies for estimating cooling timescales.

Our current findings can be summarized as follows:
\begin{itemize}
\item Power spectrum of EX Hya at Fourier frequencies above 0.05 Hz can be adequately described by a power law $P(f)\propto f^{-\alpha}$ with a slope $\alpha\sim 2.4$.
\item With the available data set we place an upper limit on the hot plasma cooling timescale on the order of $0.3-0.5$ seconds.
\item  This hot plasma cooling time puts an upper limit on the area of the accretion curtain footprints on WD surface of EX Hya, $A<10^{15}$ cm$^2$, and a lower limit on the mass flux of $\dot{M}/A>3$ g sec$^{-1}$ cm$^{-2}$.
\item Our findings imply that the electron scattering (Compton) optical depth of the hot post-shock plasma in the vertical direction might be close to, or more, than unity, which would lead to some distortions of the emergent energy spectrum and its angular distribution.
\item We deduce that the density of hot plasma in accretion curtains should be in the range $n\sim 10^{16-17}$ cm$^{-3}$. This range of densities can be probed with X-ray iron emission line diagnostics with future high energy resolution X-ray calorimeters.
\item We conclude that our estimate of the accretion curtain footprint area is significantly smaller than deduced from the eclipse mapping technique. This happens because eclipse mapping effectively determines the maximum geometrical sizes of curtains and thus their maximal possible areas within those size limits.
\item From the thickness of the accretion curtain at the WD surface, we determine that the extent of the coupling region at the boundary of WD magnetosphere in EX Hya to be $\Delta r/r<10^{-3}$, which is in line with presently available theoretical and numerical estimates of this value.
\item If we assume that coupling region at the boundary of magnetosphere has a similar value $\Delta r/r<10^{-3}$ in the case of accreting neutron stars, then we can estimate the fractional area of the neutron star accretion column $A/4\pi r_{\rm NS}^2<10^{-5}$. Such a small fractional area of the column should result in very high mass flux and thus should have strong influence on the column structure.
\end{itemize}

\section{Acknowledgements}
This material is based upon work supported financially by the
National Research Foundation. Any opinions, findings and conclusions
or recommendations expressed in this material are those of
the author(s) and therefore the NRF does not accept any liability in
regard to thereto.
This work was partially supported by grants of Presidium of Russian Academy of Sciences P21, program OFN16 of RAS, by grants NSH-5603.2012.2, RFBR-13-02-00741, RFBR-14-02-93965 as well as grants from the South African National Research Foundation (NRF) Professional Development Programme (PDP).

Some of the observations reported in this paper were obtained with the Southern African Large Telescope (SALT).

\end{document}